\documentclass[prd,aps,preprintnumbers,nofootinbib,twocolumn]{revtex4}
\usepackage{epsfig}
\usepackage{amsmath}
\usepackage{hyperref}

\def\t{{\bar t}}
\def\Mtt{m_{t\bar{t}}}

\def\Ytt{y_{t\bar{t}}}
\def\PTtt{p_{T, t\bar{t}}}
\def\Bztt{\beta_{z, t\bar{t}}}
\def\as{\alpha_S}
\def\AC{$A_{\rm C}$}
\def\AFB{$A_{\rm FB}$}
\def\dmody{\Delta|y|}

\begin{document}

\preprint{Cavendish-HEP-17/13, CP3-17-46, NIKHEF/2017-59, TUM-HEP-1109/17, TTK-17-37}

\renewcommand{\thefigure}{\arabic{figure}}

\title{The top-quark charge asymmetry at the LHC and Tevatron\\ through NNLO QCD and NLO EW}

\author{Micha\l{}  Czakon}
\affiliation{{\small Institut f\"ur Theoretische Teilchenphysik und Kosmologie, RWTH Aachen University, D-52056 Aachen, Germany}}
\author{David Heymes}
\author{Alexander Mitov}
\affiliation{{\small Cavendish Laboratory, University of Cambridge, Cambridge CB3 0HE, UK}}
\author{Davide Pagani}
\affiliation{{\small Technische Universit\"{a}t M\"{u}nchen, James-Franck-Str.~1, D-85748 Garching, Germany}}
\author{Ioannis Tsinikos}
\affiliation{{\small Centre for Cosmology, Particle Physics and Phenomenology (CP3), Universit\'e Catholique de Louvain}}
\affiliation{{\small Technische Universit\"{a}t M\"{u}nchen, James-Franck-Str.~1, D-85748 Garching, Germany}}
\author{Marco Zaro}
\affiliation{{\small Nikhef, Science Park 105, NL-1098 XG Amsterdam, The Netherlands}}
\affiliation{{\small Sorbonne Universit\'es, UPMC Univ. Paris 06, UMR 7589, LPTHE, F-75005, Paris, France}}
\affiliation{{\small CNRS, UMR 7589, LPTHE, F-75005, Paris, France}}

\date{\today}

\begin{abstract}
We provide the most up-to-date predictions for the top-quark charge asymmetry \AC{} at the LHC with a centre-of-mass energy of 8 TeV. Our result is accurate at the NNLO level in the strong interactions and at the complete-NLO level in both the strong and electroweak interactions. We present results for the inclusive \AC{}, several differential asymmetries as well as the $\PTtt$ and $\Mtt$ cumulative asymmetries. Similarly to the Tevatron forward-backward asymmetry \AFB, both NNLO QCD and NLO EW corrections to \AC{} are found to be significant. The inclusion of higher-order corrections reduces the residual scale dependence in the predicted \AC{}. The pattern of higher-order corrections indicates good perturbative control over these observables. We conclude that at present there is nearly uniform agreement between Standard Model predictions and LHC measurements across all \AC-related observables. All previously published differential \AFB{} predictions at the Tevatron are updated to the same accuracy.
\end{abstract}
\maketitle

\section{Introduction}

The charge asymmetry in top-quark pair production at hadron colliders has attracted a lot of interest in the last decade. To a large degree this interest was driven by Tevatron measurements \cite{Aaltonen:2012it,CDF:2013gna,Abazov:2014cca,Aaltonen:2017efp}, whose deviation from the predictions of the Standard Model (SM) \cite{Kuhn:1998jr,Kuhn:1998kw,Antunano:2007da,Dittmaier:2007wz,Dittmaier:2008uj,Manohar:2012rs,Almeida:2008ug,Bernreuther:2010ny,Melnikov:2010iu,Kidonakis:2011zn,Ahrens:2011uf,Hollik:2011ps,Kuhn:2011ri,Campbell:2012uf,Brodsky:2012ik,Skands:2012mm,Bernreuther:2012sx,Aguilar-Saavedra:2014kpa,Czakon:2014xsa,Czakon:2016ckf} were often interpreted as a sign of possible BSM physics. The uncertainty of the Tevatron measurements of the forward-backward asymmetry \AFB{} is not small and its main source is the limited number of top quark events. Given that the LHC has already accumulated enormous statistics of top-pair events, one can try to measure the related central-peripheral charge asymmetry \AC{} and look for a deviation from the SM in a way that is independent from the Tevatron measurements.

The measurement of \AC{} at the LHC, however, comes with its own challenges. The LHC \AC{} is very small, about 1\%, \cite{Kuhn:2011ri,Bernreuther:2012sx} which is about 10 times smaller than the Tevatron asymmetry. Such a difference between the asymmetries at these two colliders is mainly due to the fact that top-pair production at the Tevatron is driven by $q{\bar q}$ initial states, which generate a charge asymmetry, while at the LHC top pairs are mainly produced in $gg$ initial states, which are charge symmetric.

The goal of this work is to provide predictions for both the inclusive and differential \AC{} at the LHC by fully accounting for NNLO QCD corrections at $\mathcal O(\alpha_s^4)$ as well as for the
complete-NLO ones. The term complete-NLO was recently introduced in refs.~\cite{Frederix:2016ost, Biedermann:2017bss, Czakon:2017wor, Frederix:2017wme} and, unlike NLO EW, denotes the complete set of SM corrections. For $t \bar t$ production this corresponds to the NLO QCD at $\mathcal O(\alpha_s^3)$, the NLO electro-weak (EW) at $\mathcal O(\alpha \as^2)$ as well as the contributions at $\mathcal O(\alpha^2 \as)$ and $\mathcal O(\alpha^3)$ together with the LO ones at $\mathcal O(\alpha \as)$ and  $\mathcal O(\alpha^2)$.  As was established at the Tevatron effects of both QCD and EW origin are numerically significant and have to be included in any realistic comparison between LHC measurements and SM theory. We also use this opportunity to update earlier NNLO QCD differential Tevatron \AFB{} predictions \cite{Czakon:2014xsa,Czakon:2016ckf} by merging them consistently with the complete-NLO predictions.

\section{Setup of the calculation}

We calculate the top-quark charge asymmetry \AC{} at the LHC both at the inclusive level and differentially with respect to the following kinematic variables: the $t\t$ system's mass $\Mtt$, rapidity $\Ytt$, transverse momentum $\PTtt$ and longitudinal boost along the $z$-axis:
\begin{equation}
\Bztt = \frac{p_{z,t} + p_{z,\t}}{E_t + E_\t}\,,
\label{eq:Bztt}
\end{equation}
where $p_z$ and $E$ are the longitudinal momentum and energy of the top/antitop quark in the detector frame, respectively. In the notation of refs.~\cite{Czakon:2014xsa,Czakon:2016ckf} and following the setup of the CMS and ATLAS collaborations \cite{Khachatryan:2015oga,Aad:2015noh} the differential \AC{} reads:
\begin{equation}
A_{\rm C} = \frac{\sigma^+_{\rm bin} - \sigma^-_{\rm bin}}{\sigma^+_{\rm bin} + \sigma^-_{\rm bin}}~,~~ \sigma^\pm_{\rm bin} = \int \theta(\pm\dmody)\theta_{\rm bin} d\sigma\,,
\label{eq:AC}
\end{equation}
where $\dmody = |y_t| - |y_\t|$ and $d\sigma$ is the fully differential $t\t$ cross section evaluated to the required order in perturbation theory. The binning function $\theta_{\rm bin}$ takes values zero or unity and restricts to a given bin the kinematics of the $t\t$ pair in one of the four kinematic variables specified above. The inclusive asymmetry is obtained by setting $\theta_{\rm bin}=1$. 

The Tevatron forward-backward asymmetry \AFB{} is defined similarly to \AC:
\begin{equation}
A_{\rm FB} = \frac{\hat\sigma^+_{\rm bin} - \hat\sigma^-_{\rm bin}}{\hat\sigma^+_{\rm bin} + \hat\sigma^-_{\rm bin}}~,~~ \hat\sigma^\pm_{\rm bin} = \int \theta(\pm \Delta y)\theta_{\rm bin} d\sigma\,,
\label{eq:AFB}
\end{equation}
with $\Delta y = y_t - y_\t$. 

The calculation of \AC{} at NNLO QCD + NLO EW accuracy follows ref.~\cite{Czakon:2017wor} where full details about the merging of QCD and EW corrections can be found. In this paper we will use ``NLO EW'' as well ``EW'' as a short-hand notation to refer to all non-pure-QCD corrections. The genuine NLO EW correction at $\mathcal O(\alpha \as^2)$ is the dominant one among them.

The predictions in ref.~\cite{Czakon:2017wor} were obtained by combining NNLO QCD and NLO EW corrections in the so-called multiplicative approach. In this work, however, all predictions for the charge asymmetry are based on the additive approach, and the multiplicative approach will not be employed. The reason is that, as discussed in ref.~\cite{Czakon:2017wor}, the multiplicative approach is valid when QCD and EW corrections are dominated by soft and weak-Sudakov effects, respectively, since both these effects factorize. In the case of the asymmetry, neither of these two conditions are fulfilled as QCD corrections receive important contributions from hard radiation~\cite{Czakon:2014xsa} and EW corrections are dominated by QED effects~\cite{Hollik:2011ps}. Furthermore, the multiplicative combination makes use of the NLO/LO $K$-factor, which is not defined for the asymmetry. 

In the calculation of \AC{} at the LHC we use the {\sc\small LUXqed\_plus\_PDF4LHC15\_nnlo\_100} PDF set, which is based on the {\sc \small PDF4LHC} set \cite{Butterworth:2015oua, Ball:2014uwa, Harland-Lang:2014zoa, Dulat:2015mca} and employs the procedure of refs.~\cite{Manohar:2016nzj,Manohar:2017eqh} to derive the photon distribution, as well as the dynamic scale of ref.~\cite{Czakon:2016dgf}:
\begin{equation}
\mu_{F,R} = \frac{H_T}{4} = \frac{1}{4} \left( m_{T,t} + m_{T,\bar t} \right)\,,
\label{eq:scaleHT}
\end{equation}
already utilized in the differential predictions of ref.~\cite{Czakon:2017wor}. All results are for $m_t=173.3$ GeV and LHC with c.m. energy of 8 TeV. The NNLO QCD corrections to the differential observables are computed following the approach of refs.~\cite{Czakon:2015owf,Czakon:2016dgf} while the complete-NLO predictions are computed within a private extension of the {\sc\small MadGraph5\_aMC@NLO} code \cite{Alwall:2014hca,Frixione:2015zaa,Pagani:2016caq} that has already been successfully employed for the computations at complete-NLO accuracy in refs.~\cite{Frederix:2016ost, Frederix:2017wme}. We have checked that with our setup we can reproduce the results in refs.~\cite{Hollik:2011ps,Bernreuther:2012sx} for the asymmetry at NLO QCD + EW accuracy. 

The calculation of the asymmetry \AC{} is complicated by the presence of large symmetric components in the cross sections $\sigma^\pm$ defined in eq.~(\ref{eq:AC}). In order to be able to evaluate \AC{} with sufficient numerical precision, especially in NNLO QCD, we have excluded the contributions with $gg$ initial state and in general all the Born-QCD configurations. This way at NLO QCD and NLO EW we are able to evaluate \AC{} differentially with a very small Monte Carlo (MC) integration error (typically well below 1\% in each bin). The MC error of the NNLO QCD calculation is larger and in few bins it becomes comparable to the scale uncertainty. To account for the sizable MC error we include it consistently in all our predictions by adding the MC and scale errors in quadrature (independently for the upper and lower scale variations). Since the PDF dependence of the asymmetry is known to be very small compared to
the scale uncertainty, we do not quote PDF errors in our predictions.

All differential asymmetries are ratios of ``unexpanded'' numerators and denominators, {\it i.e.}, both the numerator and denominator in each bin are evaluated through the corresponding order in perturbation theory and then the ratio is taken. The effect of the series expansion of the ratio will be shown in the next section only for the inclusive asymmetry \AC{}.

Unlike the LHC calculations discussed above, the updated Tevatron \AFB{} prediction is derived with the setup used in refs.~\cite{Czakon:2014xsa,Czakon:2016ckf}, namely, with fixed renormalization and factorization scales $\mu_{F,R}=m_t$ and MSTW2008 \cite{Martin:2009iq} PDF sets. This updated calculation has been utilized in the latest Tevatron combination ref.~\cite{Aaltonen:2017efp}.

\section{Predictions for \AC{} at the LHC}

\subsection{Inclusive asymmetry}
\label{sec:inclusiveAC}

\begin{figure}[t]
  \centering
     \hspace{-1mm} 
     \includegraphics[width=3.5in]{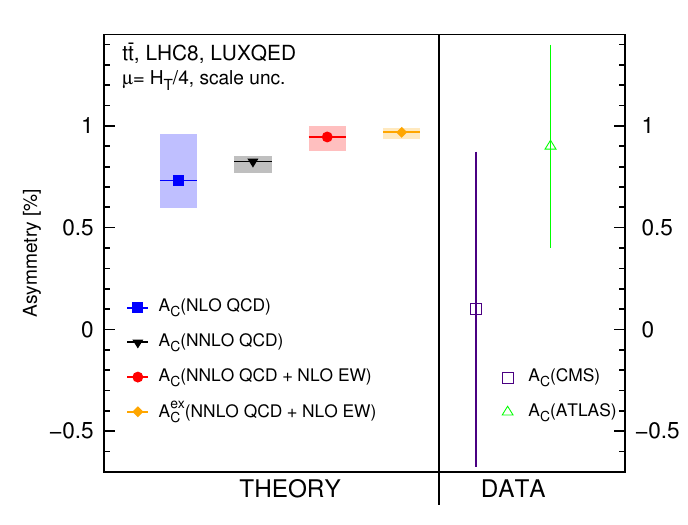} 
   \caption{Inclusive charge asymmetry \AC{} for the LHC at 8 TeV in NLO QCD, NNLO QCD and NNLO QCD + NLO EW versus CMS and ATLAS measurements \cite{Aad:2015noh,Khachatryan:2015oga}.}
   \label{fig:inclusive}
\end{figure}

In tables~\ref{tab:AC-HT4} and \ref{tab:AC-mt} and in fig.~\ref{fig:inclusive} we show the predictions for the inclusive \AC{} for the LHC at 8 TeV in NLO QCD, NLO QCD + NLO EW, NNLO QCD and in NNLO QCD + NLO EW. By default we calculate the inclusive \AC{} based on the unexpanded definition eq.~(\ref{eq:AC}) and using the dynamic scale eq.~(\ref{eq:scaleHT}). As mentioned in the Introduction, and in line with the previous literature, we also show predictions for the expanded version $A^{\rm ex}_{\rm C}$ of the inclusive \AC{}.

Defining the expanded asymmetry $A^{\rm ex}_{\rm C}$ is not straightforward, especially in the presence of EW corrections since two different coupling constants are present. In pure QCD the expanded version is traditionally defined through the expansion of the perturbative cross sections in the ratio (\ref{eq:AC}). Such an approach leads to ambiguity related to the order of the partonic distributions which contain implicit $\as$ dependence. Once EW corrections are added such an expansion becomes even more cumbersome, especially at high perturbative orders. 

In view of the above mentioned difficulties we introduce $A^{\rm ex}_{\rm C}$ through the following simplified definition:
\begin{eqnarray}
A^{\rm ex,(1)}_{\rm C} &=& A^{(1)}_{\rm C} K^{(1)}\,, \label{eq:expand-nlo}\\
A^{\rm ex,(2)}_{\rm C} &=& A^{(2)}_{\rm C} K^{(2)}  - A^{(1)}_{\rm C} (K^{(1)} - 1)K^{(1)}\,, \label{eq:expand-nnlo}
\end{eqnarray}
which we use in both pure QCD through NNLO and in the presence of EW corrections through NNLO QCD + NLO EW.

In eqns.~(\ref{eq:expand-nlo},\ref{eq:expand-nnlo}) above $A^{(n)}_{\rm C}$ is the N$^n$LO QCD approximation for \AC{} (with $A^{(0)}_{\rm C}=0$)
\begin{equation}
A^{(n)}_{\rm C} = \frac{\as^3N_3+\dots +\as^{n+2}N_{n+2}}{\as^2D_2+\dots +\as^{n+2}D_{n+2}}\,,
\label{eq:An}
\end{equation}
with $D_n$ and $N_n$ originating, respectively, in the perturbative expansion of the denominator and numerator in eq.~(\ref{eq:AC}). The $K$-factor $K^{(n)}$ is the ratio of the inclusive N$^n$LO and LO $t\t$ cross sections, which is always computed in pure QCD 
\begin{equation}
K^{(n)} = \frac{\as^2D_2+\dots +\as^{n+2}D_{n+2}}{\as^2D_2}\,.
\label{eq:Kn}
\end{equation}

For consistency, we use the value of the inclusive cross section obtained by summing the bins of the differential distributions. As pointed out in ref.~\cite{Czakon:2016dgf} due to the use of dynamic scales this value is slightly different from the one obtained with the program {\tt Top++} \cite{Czakon:2011xx}.

One can easily check that in pure QCD eqns.~(\ref{eq:expand-nlo},\ref{eq:expand-nnlo}) are exactly equal to the expansion up to NLO and NNLO accuracy, respectively, and no higher-order terms are introduced. These equations can also be used to define $A^{\rm ex}_{\rm C}$ in the presence of EW corrections if their effects in the \AC{} denominator and $K$-factors are neglected. This is well justified since the size of EW corrections to the inclusive $t\t$ cross section is known to be small, about 1\%. 

Therefore, we also use eqns.~(\ref{eq:expand-nlo},\ref{eq:expand-nnlo}) to define $A^{\rm ex}_{\rm C}$ through (N)NLO QCD + NLO EW, with $A^{(n)}_{\rm C}$ computed in QCD+EW and $K^{(n)}$ restricted to pure QCD. The impact of EW corrections in $A^{\rm ex,(1)}_{\rm C}$ is enhanced with respect to $A^{(1)}_{\rm C}$ due to the factor $K^{(1)}$ in eq.~(\ref{eq:expand-nlo}). Including EW corrections only in the asymmetries is equivalent to the expanded definition for the asymmetry through NNLO QCD + NLO EW given in eq.~(3) of ref.~\cite{Czakon:2014xsa}. 

Eqns.~(\ref{eq:expand-nlo},\ref{eq:expand-nnlo}) can also be used to calculate the scale dependence of $A^{\rm ex}_{\rm C}$ by evaluating each factor on the RHS for the corresponding value of $\mu_{F,R}$.

\begin{table}[t]
\small
\renewcommand{\arraystretch}{1.25}
\begin{center}
\begin{tabular}{  c | c c c c c }
\hline\hline
& NLO QCD & NLO+EW & NNLO QCD & NNLO+EW\\
\hline
$A_{\rm C} [\%]$ & $0.73^{+0.23}_{-0.13}$ & $0.86^{+0.25}_{-0.14}$ & $0.83^{+0.03}_{-0.06}$ & $0.95^{+0.05}_{-0.07}$ \\
$A_{\rm C}^{\text{ex}} [\%]$ & $0.96^{+0.11}_{-0.09}$ & $1.13^{+0.10}_{-0.08}$ & $0.85^{+0.02}_{-0.04}$ & $0.97^{+0.02}_{-0.03}$\\
\hline
\end{tabular}
\caption{Inclusive top-quark charge asymmetry at NLO QCD, NLO QCD + NLO EW, NNLO QCD and NNLO QCD + NLO EW with $\mu_F=\mu_R=H_T/4$. Errors are from scale variation.}
\label{tab:AC-HT4}
\end{center}
\end{table}
\noindent
\begin{table}[t]
\small
\renewcommand{\arraystretch}{1.25}
\begin{center}
\begin{tabular}{  c | c c c c c }
\hline\hline
& NLO QCD & NLO+EW & NNLO QCD & NNLO+EW\\
\hline
$A_{\rm C} [\%]$ & $0.64^{+0.16}_{-0.10}$ & $0.76^{+0.17}_{-0.10}$ & $0.79^{+0.05}_{-0.06}$ & $0.90^{+0.07}_{-0.07}$ \\
$A_{\rm C}^{\text{ex}} [\%]$ & $0.94^{+0.10}_{-0.08}$ & $1.13^{+0.09}_{-0.07}$ & $0.86^{+0.02}_{-0.04}$ &  $0.95^{+0.02}_{-0.04}$  \\
\hline
\end{tabular}
\caption{As in table~\ref{tab:AC-HT4} but with fixed scales $\mu_F=\mu_R=m_t$.}
\label{tab:AC-mt}
\end{center}
\end{table}

From tables~\ref{tab:AC-HT4} and \ref{tab:AC-mt} we conclude that the effect of the NNLO QCD correction on the inclusive \AC{} at the LHC is different for the cases of the expanded and unexpanded definitions; this is at variance with the pattern observed in ref.~\cite{Czakon:2014xsa} for the Tevatron \AFB{}. As can be seen in table~\ref{tab:AC-HT4}, the NNLO QCD correction increases the asymmetry by 0.10\% in the expanded case, while in the unexpanded case it reduces the asymmetry by roughly the same amount. In both cases the inclusion of the NNLO QCD correction leads to a strong reduction of the scale uncertainty which completely overlaps with the uncertainty band of the NLO prediction. This fact is indicative of the good consistency of the pure QCD predictions. Furthermore, the difference between the expanded and unexpanded predictions is reduced at NNLO, which is another sign of the reliable theoretical control over these 
\widetext

\begin{figure}[h]
  \centering
     \hspace{-1mm} 
     \includegraphics[width=3.1in]{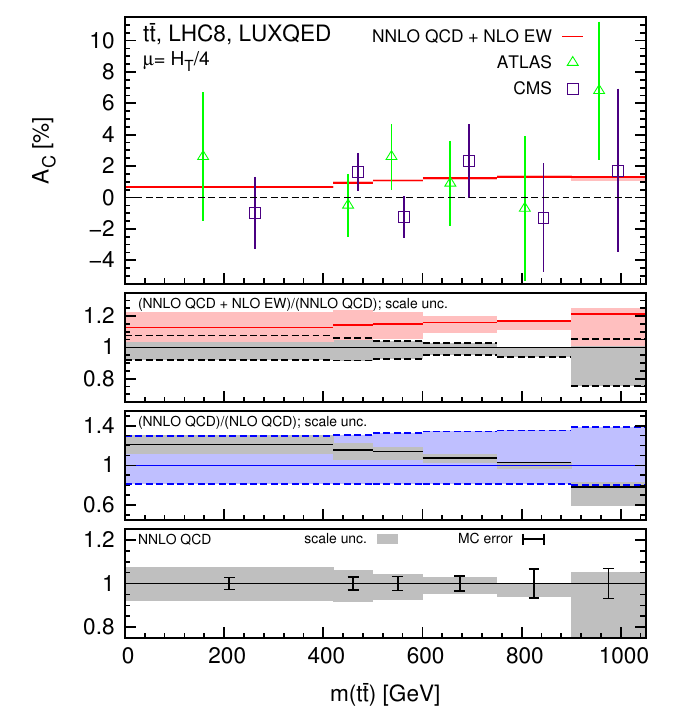}
     \includegraphics[width=3.1in]{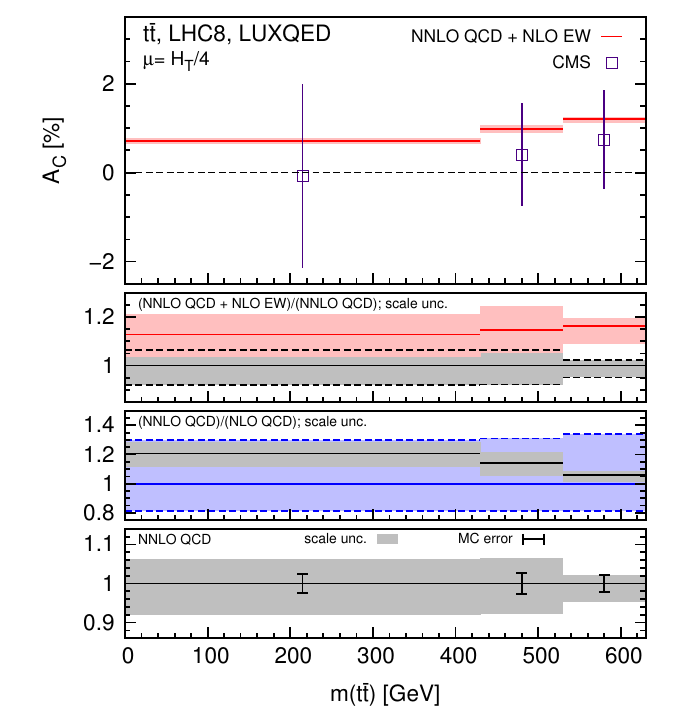}  
   \caption{The $\Mtt$-dependent differential asymmetry for two sets of bins versus measurements from ATLAS and CMS.}
   \label{fig:Mtt}
\end{figure}
\endwidetext
\hskip -3mm
predictions. 

The effect of EW corrections is always positive, and it increases the unexpanded asymmetries at NLO and NNLO QCD, as well as the expanded one at NNLO QCD, by about 0.12\%. The effect of EW corrections is larger (0.17\%) when they are combined with the expanded asymmetry at NLO QCD, because of the NLO/LO $K$-factor that enters the expanded asymmetry definition. In all cases EW corrections are by far dominated by the $\mathcal O (\alpha\as^2)$ contribution. In fig.~\ref{fig:inclusive} we compare our predictions for the asymmetry to CMS \cite{Khachatryan:2015oga} and ATLAS \cite{Aad:2015noh} measurements where a very good theory-data agreement is found. As it will also be the case for the differential asymmetries, experimental errors largely outweigh the theoretical ones.

In order to facilitate comparisons with existing calculations, in table~\ref{tab:AC-mt} we also provide predictions derived with fixed scales $\mu_F=\mu_R=m_t$. The pattern of higher-order corrections is similar to the case of dynamic scales. The NLO QCD and NLO + EW values of the expanded asymmetry $A_{\rm C}^{\text{ex}}$ from this table agree within statistical uncertainties with the numbers in table~8 of ref.~\cite{Bernreuther:2012sx} which have been obtained with a slightly different setup. The difference between the NNLO predictions with the two scale choices is reduced, as expected, with respect to the case of NLO predictions. In all cases, the two scale choices are always compatible within scale uncertainties. Comparing the results in tables~\ref{tab:AC-HT4} and \ref{tab:AC-mt} we also note the very small dependence of the expanded asymmetry $A_{\rm C}^{\text{ex}}$ on the scale choice.

\subsection{Differential asymmetry}\label{sec:diff-AC}

In this section we present predictions for several differential charge asymmetries and compare them with existing LHC measurements \cite{Aad:2015noh, Khachatryan:2015oga}. All calculations are performed in NNLO QCD + NLO EW using the unexpanded definition in eq.~(\ref{eq:AC}). The reason of this choice is twofold. First, we have verified in sec.~\ref{sec:inclusiveAC} that differences among expanded and unexpanded predictions are negligible when NNLO QCD corrections are taken into account. Second, at the differential level, the contribution of EW corrections to the $K$-factors can no longer be neglected, overcomplicating the simultaneous expansion in QCD and EW couplings.

In the main panels of fig.~\ref{fig:Mtt} we compare the $\Mtt$-dependent charge asymmetry and two available measurements. The theory uncertainty is negligible compared to the experimental one. All measurements are consistent with the theory predictions within uncertainties. 

In order to gain more insight into the structure of the theory prediction, in the first insets we compare the scale variations of the NNLO QCD result (the grey band) and of the NNLO QCD + NLO EW one (the red band), both normalized to the NNLO QCD central value. Clearly, the impact of EW corrections on the differential asymmetry is significant. The EW corrections tend to increase the $\Mtt$-dependent charge asymmetry from about 13\% close to threshold to slightly over 20\% at around 1 TeV. The inclusion of EW corrections does not noticeably affect the size of the scale uncertainty. 

In the second insets we analyze the size of the higher-order QCD corrections. Specifically, we plot as a grey 
\widetext

\begin{figure}[t]
  \centering
     \hspace{-1mm} 
     \includegraphics[width=3.1in]{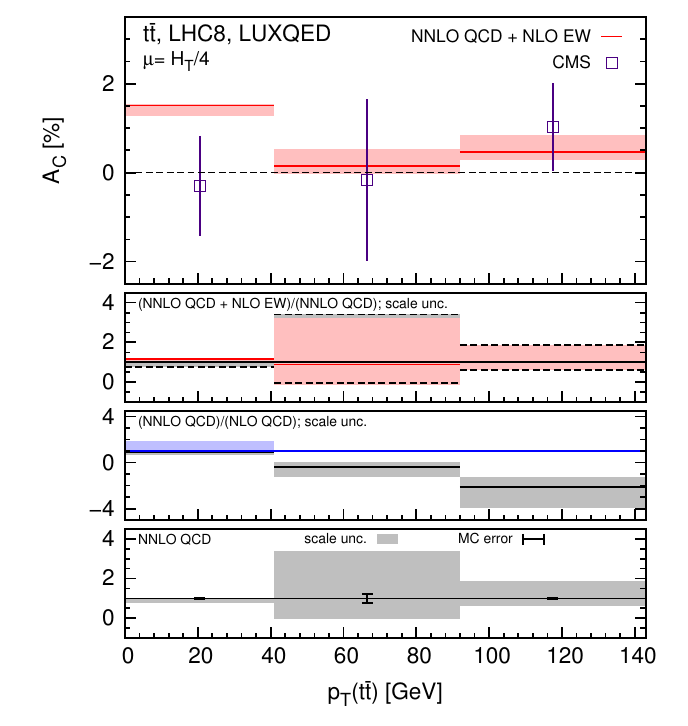} 
     \includegraphics[width=3.1in]{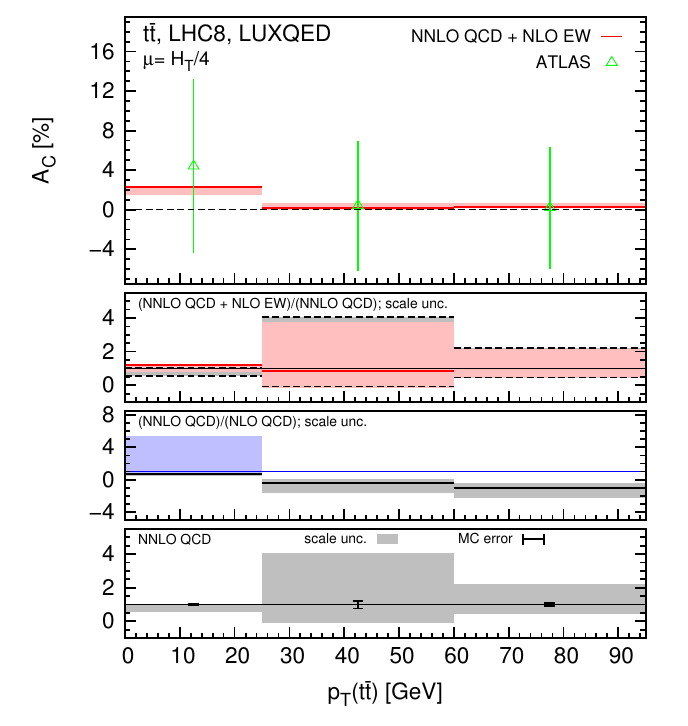} 
   \caption{As in fig.\ref{fig:Mtt} but for the $\PTtt$-dependent asymmetry.}
   \label{fig:PTtt}
\end{figure}
\begin{figure}[t]
  \centering
     \hspace{-1mm} 
     \includegraphics[width=3.1in]{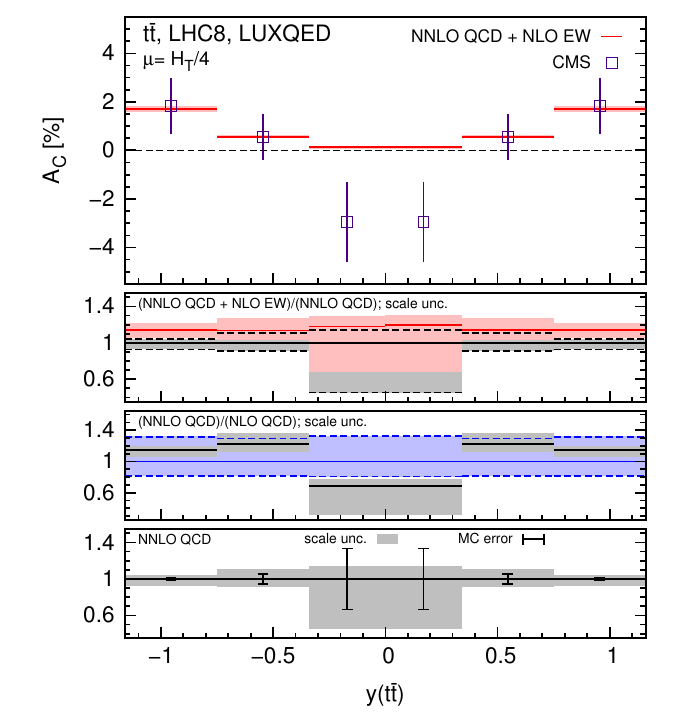}
     \includegraphics[width=3.1in]{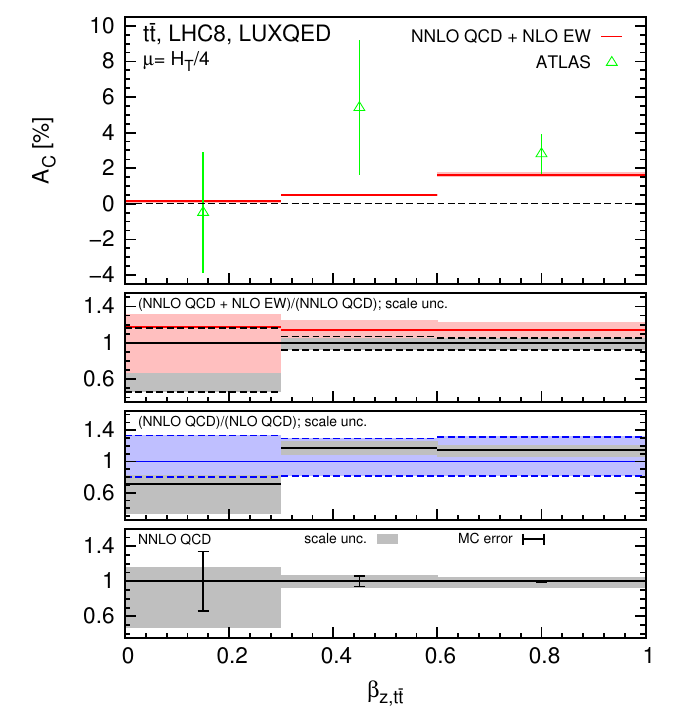}  
   \caption{The $\Ytt$-dependent differential asymmetry versus CMS measurement (left) and the $\Bztt$-dependent differential asymmetry versus ATLAS measurement (right).}
   \label{fig:Ytt-betaZtt}
\end{figure}
\endwidetext
\hskip -3.5mm
band the NNLO/NLO $K$-factor including its scale variation. We observe that the NNLO QCD correction has very significant impact on the $\Mtt$-dependent charge asymmetry. It significantly modifies the shape of the distribution by increasing it by about 20\% close to threshold and by lowering it approximately by the same amount for $\Mtt$ about 1 TeV. The scale error is reduced significantly once the NNLO correction is included. This can be seen by comparing with the blue band which represents the relative scale uncertainty at NLO QCD. 

Finally, in the third insets of fig.~\ref{fig:Mtt} we show the MC error of the NNLO result. In order to better judge the significance of the MC error we compare it with the NNLO scale error. We conclude that above about 700 GeV the MC error is as large as the scale one. Given it is significant, the theory error of our final predictions includes the MC error added in quadrature.

The $\PTtt$-dependent charge asymmetry is shown in fig.~\ref{fig:PTtt}, where it is compared to available measurements from CMS~\cite{Khachatryan:2015oga} (left) and ATLAS~\cite{Aad:2015noh} (right). When interpreting this distribution one should keep in mind that the QCD prediction is of NNLO accuracy only in the first (leftmost) bin, while in the other bins it is of at most NLO accuracy. This also
explains the large effects and scale uncertainties which are observed in the NNLO/NLO $K$-factor for the bins with $\PTtt > 0$.  Theory and data again agree, with only a mild tension (below $2\sigma$) in the first CMS bin. 

The theory errors are again significantly smaller than the experimental ones although not as much as in the $\Mtt$ case. The overall pattern of higher-order corrections is very similar to the one known for a long time from the Tevatron: one has positive higher-order corrections in the first bin and large negative corrections in the subsequent bins.  The absolute size of the EW corrections remains similar to what has been found for other observables. However, their impact is relatively small with respect to the size of the scale error which is particularly large for the second bin.  For the same reason, the MC error in this distribution is completely negligible. 

In the main panel of fig.~\ref{fig:Ytt-betaZtt}(left) we show the $\Ytt$-dependent asymmetry. We compare it with the available CMS measurement~\cite{Khachatryan:2015oga}. The theory error is again completely negligible in all bins when compared to the experimental one. In all bins but the central two data and theory agree perfectly. In the central two bins the measurement is lower than theory by less than 2$\sigma$. Given the large experimental errors, this discrepancy does not appear to be significant.

The higher-order corrections in the $\Ytt$-asymmetry have an interesting pattern. EW corrections give a contribution which is positive and nearly constant - as large as 20\% of the NNLO QCD one. The NNLO QCD correction has similar behavior for all bins but the two central ones, where it increases the NLO QCD result by about 20\%. In the two central bins the NNLO QCD correction lowers the NLO QCD one by about the same amount. For this reason, while the combined effect of NNLO QCD and NLO EW is to increase the NLO QCD non-central bins by about 40\%-50\%, in the two central bins they almost cancel each other, {\it i.e.}, the NLO QCD result is close to the NNLO QCD + NLO EW prediction. One should keep in mind, however, that the absolute size of the asymmetry in the two central bins is very small and therefore the absolute impact of these corrections is tiny. The relative MC error is large but its absolute size is very small in this distribution and thus numerical uncertainties do not play an important role here.

In fig.~\ref{fig:Ytt-betaZtt}(right) we show the last distribution studied by us, namely, the $\Bztt$-dependent asymmetry. Data and theory agree in all bins with theory being slightly outside the experimental error band in the second bin. The theory error is again negligible compared to the experimental one. We observe positive NLO EW and NNLO QCD corrections of about 20\% for all bins except the first one. In the first bin NLO EW corrections remain positive but NNLO QCD corrections are negative, as large as -30\%. The MC error is comparable to the scale error but its absolute size is small and therefore immaterial for the current theory/data comparison.

To summarize, we observe that for all differential charge asymmetries studied by us the NLO EW and NNLO QCD corrections have very significant impact on the shapes and values of these asymmetries. In general we observe a good consistency between SM theory predictions and LHC measurements for all differential asymmetries, with only mild tensions in a few bins. At present the experimental errors are much larger than the theory ones. The theory/data comparison will significantly benefit from possible future measurements with increased precision.

\subsection{Cumulative asymmetry}

In this section we give predictions for the $\PTtt$- and $\Mtt$-cumulative asymmetries. They are closely related to the corresponding differential asymmetries discussed in sec. \ref{sec:diff-AC}. The definition of the cumulative asymmetry is similar to the one in eq.~(\ref{eq:AC}): for a given value of the kinematic variable for which we compute the asymmetry, the bin ranges from zero to that value. 

While differential and cumulative asymmetries contain the same information (the former is related to the derivative of the latter), the cumulative ones tend to be better behaving since in many cases the higher-order corrections are distributed more uniformly over the full kinematic range. Cumulative asymmetries have been extensively discussed in the context of the Tevatron \AFB{}, see ref.~\cite{Czakon:2016ckf}.

\begin{figure}[t]
  \centering
     \hspace{-1mm} 
     \includegraphics[width=2.3in]{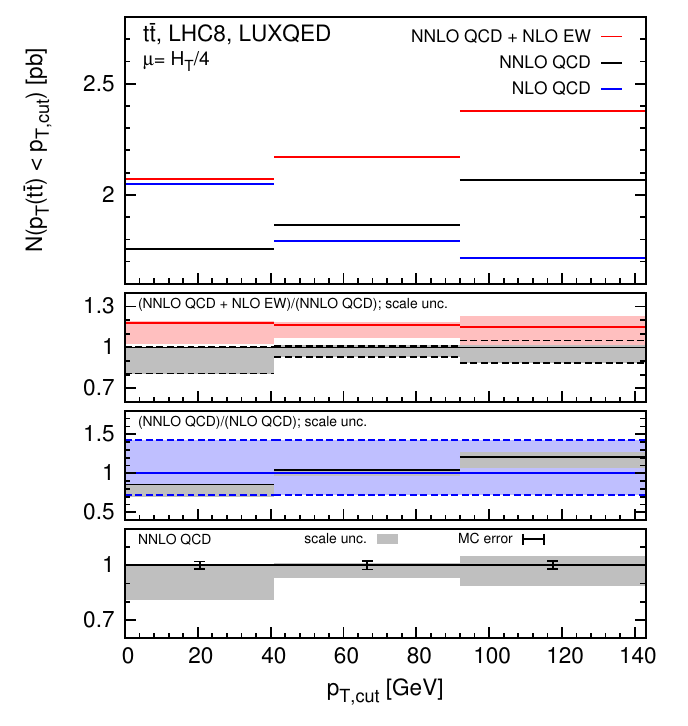}      
     \includegraphics[width=2.3in]{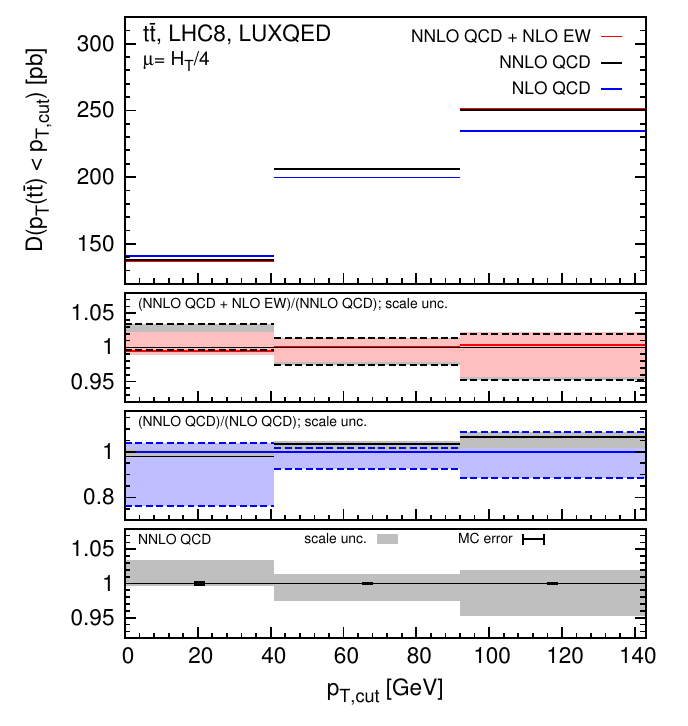}      
     \includegraphics[width=2.3in]{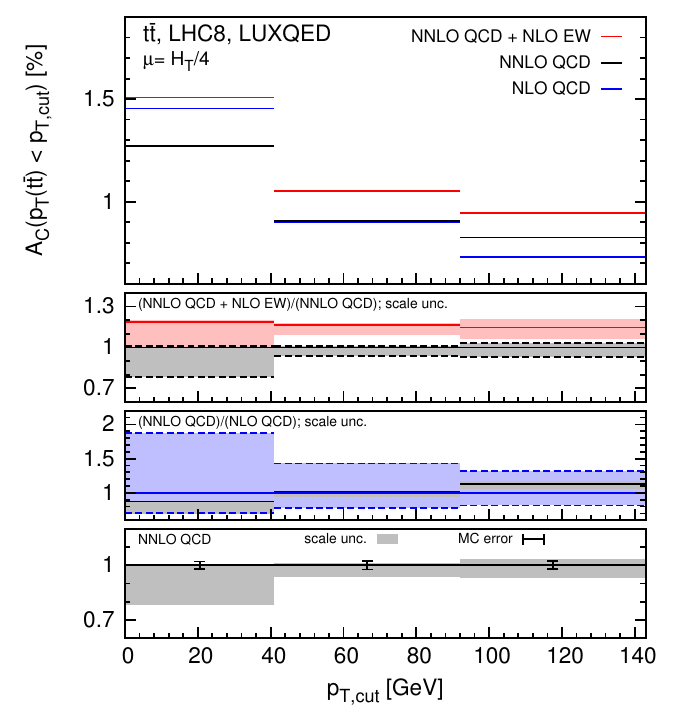}      
   \caption{The $\PTtt$-dependent cumulative numerator (left), cumulative denominator (center) and cumulative \AC{} (right).}
   \label{fig:PTtt-cumul}
\end{figure}

In fig.~\ref{fig:PTtt-cumul} we present the predictions for the $\PTtt$-dependent cumulative \AC{}. We present separately the numerator, denominator (both defined in the sense of eq.~(\ref{eq:AC})) and the complete asymmetry. The MC error in all bins is small. Comparing to fig.~\ref{fig:PTtt} we see that the NNLO/NLO $K$-factor for the cumulative asymmetry is much smaller than the one for the differential asymmetry. Moreover, the relative scale uncertainty is also much smaller for the cumulative distribution than for the differential one.

The significance of the $\PTtt$ cumulative asymmetry has been extensively discussed in refs.~\cite{Czakon:2014xsa,Czakon:2016ckf}. It allows one to disentangle the contributions from soft and hard emissions to \AC{} and thus compare fixed-order predictions with ones based on soft-gluon resummation. Although the bins used in this work are much wider than the bins used in the Tevatron analysis of refs.~\cite{Czakon:2014xsa,Czakon:2016ckf}, one can immediately see that \AFB{} and \AC{} behave similarly. The effect of the higher-order EW and QCD corrections is again very significant and each of them leads to a significant modification to the shape of the asymmetric numerator. As expected, the effect of NLO EW corrections on the (symmetric) denominator is completely negligible 
\widetext

\endwidetext
\begin{figure}[t]
  \centering
     \hspace{-1mm} 
     \includegraphics[width=3.5in]{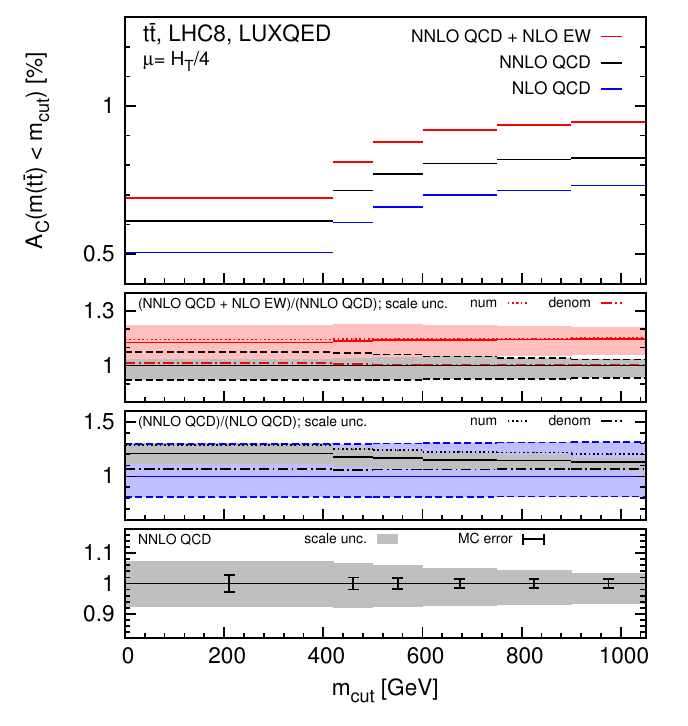}
   \caption{$\Mtt$-dependent cumulative asymmetry in NNLO QCD + NLO EW.}
   \label{fig:Mtt-cumul}
\end{figure}
\hskip -3.5mm
while the effect of the NNLO QCD correction is small to moderate. The individual impact of NNLO QCD and NLO EW corrections on the $\PTtt$-dependent cumulative asymmetry is very significant and for this reason it would be very interesting to compare these predictions to data. In particular, in the 
second bin shown in the histograms in fig.~\ref{fig:PTtt-cumul}, the effect of NLO EW corrections is larger than the NNLO QCD scale uncertainty for both the cumulative asymmetry and numerator. 

Our prediction for the $\Mtt$-dependent cumulative asymmetry is given in fig.~\ref{fig:Mtt-cumul}. Unlike the corresponding differential case, fig.~\ref{fig:Mtt}, the MC error of the cumulative distribution is small in all bins. Also, the NNLO/NLO $K$-factor is much flatter, which means that the higher-order corrections to the shape are smaller than in the differential case. We note that the current calculation is defined differently than the one in ref.~\cite{Czakon:2016ckf} with the definition in ref.~\cite{Czakon:2016ckf} being $A_C (\Mtt > m_{\rm cut})$.

\section{Updated Tevatron \AFB{} predictions}

In this section we update the Tevatron predictions of refs.~\cite{Czakon:2014xsa,Czakon:2016ckf} by consistently merging QCD and EW corrections in the differential asymmetry. With the exception of the inclusive \AFB{}, the previous NNLO QCD predictions \cite{Czakon:2014xsa,Czakon:2016ckf} for the differential \AFB{} did not include any EW corrections.

\begin{figure}[h]
  \centering
     \hspace{-1mm} 
     \includegraphics[width=2.3in]{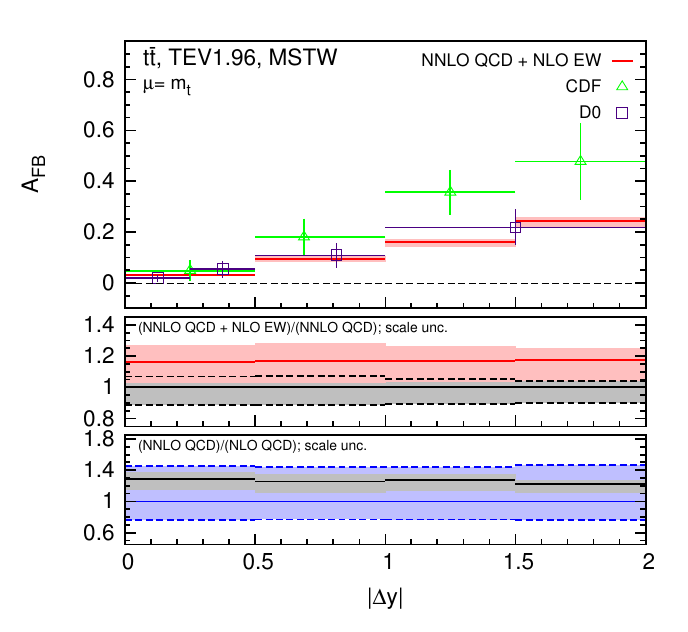}
     \includegraphics[width=2.3in]{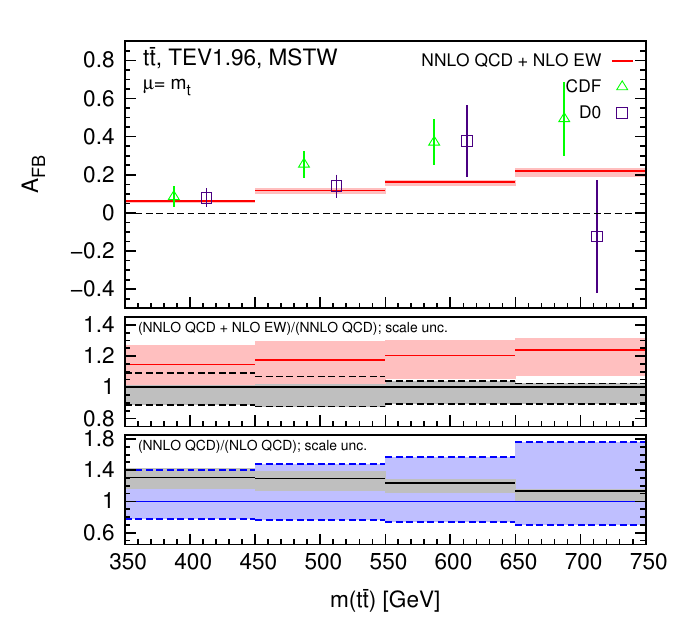}
     \includegraphics[width=2.3in]{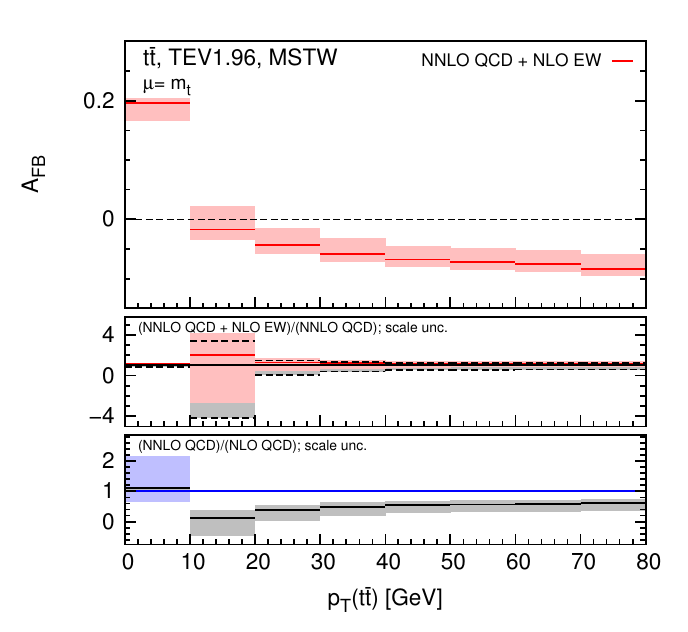}
   \caption{Differential $|\Delta y|$, $\Mtt$ and $\PTtt$ \AFB{} asymmetries at the Tevatron in NNLO QCD + NLO EW versus available data from the CDF~\cite{Aaltonen:2012it} and D\O ~\cite{Abazov:2014cca} collaborations.}
   \label{fig:AFB-Tev}
\end{figure}

The updated results are shown in fig.~\ref{fig:AFB-Tev}. We consider the $|\Delta y|$, $\Mtt$ and $\PTtt$ differential \AFB{}. The NLO and NNLO QCD results are the same as the ones in refs.~\cite{Czakon:2014xsa,Czakon:2016ckf}. For consistency with these previously published results here we use the same settings, in particular, the same PDF sets ({\sc\small MSTW2008} \cite{Martin:2009iq} at 68\% cl) and fixed factorization and renormalization scale $\mu_F=\mu_R=m_t$. 

The slope of the $\Mtt$-dependent \AFB{} prediction derived in this work has been extracted in ref.~\cite{Aaltonen:2017efp} where it is compared with the latest Tevatron \AFB{} combination.

The effect of the EW corrections, relative to NNLO QCD, is significant. They increase the differential $|\Delta y|$ and $\Mtt$ asymmetries by about 20\% across the kinematic range considered here. The EW corrections also slightly increase the size of the scale variation for those two observables. In the case of the $\PTtt$-dependent \AFB{} the EW corrections tend to increase the absolute size of the asymmetry by about 20\%; the only exception is the (10-20) GeV bin where the relative increase is about 100\%. This increment is not induced by the EW corrections 
\widetext

\endwidetext
\hskip -3.5mm
themselves but rather by the NNLO QCD correction which diminishes the NLO QCD prediction (see the second inset) and, in turn, increases the ratio (NNLO QCD + NLO EW)/(NNLO QCD).

\section{Summary}

In this work we calculate the top-quark charge asymmetry at a $pp$ collider like the LHC. We systematically account for NNLO QCD and complete-NLO corrections.  Overall, we find that both the NNLO QCD and EW corrections have a large impact on the predicted charge asymmetry, both inclusively and differentially. We provide a set of predictions for cumulative asymmetries which have not been measured so far but are better behaved with respect to higher-order corrections.
 
We have emphasized the analogy of \AC{} with the related \AFB{} asymmetry at the Tevatron. Indeed, we find that in both cases higher-order corrections to related differential asymmetries seem to follow very similar patterns. We also find some differences. The pattern of higher-order QCD corrections in the inclusive asymmetry is somewhat different from the \AFB{} asymmetry at the Tevatron: while the NNLO QCD corrections increase the unexpanded \AC{} relative to NLO QCD, they decrease $A^{\rm ex}_{\rm C}$ when the expanded definition is used.

Our predictions for the LHC at 8 TeV are compared with existing measurements form ATLAS and CMS. We observe very good agreement between theory and data. In few bins we observe deviations which do not appear to be significant and in all cases are below the level of $2\sigma$. The differential and inclusive LHC charge asymmetries derived in this work have been included in the latest LHC \AC{} combination \cite{Sirunyan:2017lvd}.

At present the experimental uncertainties dominate over the theory ones. Clearly, any future improvement in the measurements will be very beneficial as it will allow more detailed scrutiny of this important observable.

All results derived in this work are available in electronic form \cite{web-based-results}.

\begin{acknowledgments}
A.M. thanks the Department of Physics at Princeton University for hospitality during the completion of this work. The work of M.C. is supported in part by grants of the DFG and BMBF. The work of D.H. and A.M. is supported by the UK STFC grants ST/L002760/1 and ST/K004883/1 and by the European Research Council Consolidator Grant ``NNLOforLHC2". The work of D.P is supported by the Alexander von Humboldt Foundation, in the framework of the Sofja Kovalevskaja Award Project ``Event Simulation for the Large Hadron Collider at High Precision''. The work of I.T. is supported by the F.R.S.-FNRS ``Fonds de la Recherche Scientifique'' (Belgium). The work of M.Z. has been supported by the Netherlands
National Organisation for Scientific Research (NWO), by the European Union's Horizon 2020 research and
innovation programme under the Marie Sklodovska-Curie grant
agreement No 660171 and in part by the ILP LABEX (ANR-10-LABX-63),
in turn supported by French state funds managed by the ANR
within the ``Investissements d'Avenir'' programme
under reference ANR-11-IDEX-0004-02.
\end{acknowledgments}


\begin{thebibliography}{99}

\bibitem{Aaltonen:2012it}
  T.~Aaltonen {\it et al.} [CDF Collaboration],
  Phys.\ Rev.\ D {\bf 87} (2013) no.9,  092002
  [arXiv:1211.1003 [hep-ex]].

\bibitem{CDF:2013gna}
  T.~Aaltonen {\it et al.} [CDF Collaboration],
  Phys.\ Rev.\ Lett.\  {\bf 111} (2013) no.18,  182002
  [arXiv:1306.2357 [hep-ex]].

\bibitem{Abazov:2014cca}
  V.~M.~Abazov {\it et al.} [D0 Collaboration],
  Phys.\ Rev.\ D {\bf 90} (2014) 072011
  [arXiv:1405.0421 [hep-ex]].

\bibitem{Aaltonen:2017efp}
  T.~A.~Aaltonen {\it et al.} [CDF and D0 Collaborations],
  [arXiv:1709.04894 [hep-ex]].

\bibitem{Kuhn:1998jr}
  J.~H.~Kuhn and G.~Rodrigo,
  Phys.\ Rev.\ Lett.\  {\bf 81} (1998) 49
  [hep-ph/9802268].

\bibitem{Kuhn:1998kw}
  J.~H.~Kuhn and G.~Rodrigo,
  Phys.\ Rev.\ D {\bf 59} (1999) 054017
  [hep-ph/9807420].

\bibitem{Antunano:2007da}
  O.~Antunano, J.~H.~Kuhn and G.~Rodrigo,
  Phys.\ Rev.\ D {\bf 77} (2008) 014003
  [arXiv:0709.1652 [hep-ph]].

\bibitem{Dittmaier:2007wz}
  S.~Dittmaier, P.~Uwer and S.~Weinzierl,
  Phys.\ Rev.\ Lett.\  {\bf 98} (2007) 262002
  [hep-ph/0703120 [HEP-PH]].

\bibitem{Dittmaier:2008uj}
  S.~Dittmaier, P.~Uwer and S.~Weinzierl,
  Eur.\ Phys.\ J.\ C {\bf 59} (2009) 625
  [arXiv:0810.0452 [hep-ph]].

\bibitem{Manohar:2012rs}
  A.~V.~Manohar and M.~Trott,
  Phys.\ Lett.\ B {\bf 711} (2012) 313
  [arXiv:1201.3926 [hep-ph]].

\bibitem{Almeida:2008ug}
  L.~G.~Almeida, G.~F.~Sterman and W.~Vogelsang,
  Phys.\ Rev.\ D {\bf 78} (2008) 014008
  [arXiv:0805.1885 [hep-ph]].

\bibitem{Bernreuther:2010ny}
  W.~Bernreuther and Z.~G.~Si,
  Nucl.\ Phys.\ B {\bf 837} (2010) 90
  [arXiv:1003.3926 [hep-ph]].

\bibitem{Melnikov:2010iu}
  K.~Melnikov and M.~Schulze,
  Nucl.\ Phys.\ B {\bf 840} (2010) 129
  [arXiv:1004.3284 [hep-ph]].

\bibitem{Kidonakis:2011zn}
  N.~Kidonakis,
  Phys.\ Rev.\ D {\bf 84} (2011) 011504
  [arXiv:1105.5167 [hep-ph]].

\bibitem{Ahrens:2011uf}
  V.~Ahrens, A.~Ferroglia, M.~Neubert, B.~D.~Pecjak and L.~L.~Yang,
  Phys.\ Rev.\ D {\bf 84} (2011) 074004
  [arXiv:1106.6051 [hep-ph]].

\bibitem{Hollik:2011ps}
  W.~Hollik and D.~Pagani,
  Phys.\ Rev.\ D {\bf 84} (2011) 093003
  [arXiv:1107.2606 [hep-ph]].

\bibitem{Kuhn:2011ri}
  J.~H.~Kuhn and G.~Rodrigo,
  JHEP {\bf 1201} (2012) 063
  [arXiv:1109.6830 [hep-ph]].

\bibitem{Campbell:2012uf}
  J.~M.~Campbell and R.~K.~Ellis,
  J.\ Phys.\ G {\bf 42} (2015) no.1,  015005
  [arXiv:1204.1513 [hep-ph]].

\bibitem{Brodsky:2012ik}
  S.~J.~Brodsky and X.~G.~Wu,
  Phys.\ Rev.\ D {\bf 85} (2012) 114040
  [arXiv:1205.1232 [hep-ph]].

\bibitem{Skands:2012mm}
  P.~Skands, B.~Webber and J.~Winter,
  JHEP {\bf 1207} (2012) 151
  [arXiv:1205.1466 [hep-ph]].

\bibitem{Bernreuther:2012sx}
  W.~Bernreuther and Z.~G.~Si,
  Phys.\ Rev.\ D {\bf 86} (2012) 034026
  [arXiv:1205.6580 [hep-ph]].

\bibitem{Aguilar-Saavedra:2014kpa}
  J.~A.~Aguilar-Saavedra, D.~Amidei, A.~Juste and M.~Perez-Victoria,
  Rev.\ Mod.\ Phys.\  {\bf 87} (2015) 421
  [arXiv:1406.1798 [hep-ph]].

\bibitem{Czakon:2014xsa}
  M.~Czakon, P.~Fiedler and A.~Mitov,
  Phys.\ Rev.\ Lett.\  {\bf 115} (2015) no.5,  052001
  [arXiv:1411.3007 [hep-ph]].

\bibitem{Czakon:2016ckf}
  M.~Czakon, P.~Fiedler, D.~Heymes and A.~Mitov,
  JHEP {\bf 1605} (2016) 034
  [arXiv:1601.05375 [hep-ph]].
  
\bibitem{Frederix:2016ost}
  R.~Frederix, S.~Frixione, V.~Hirschi, D.~Pagani, H.~S.~Shao and M.~Zaro,
  JHEP {\bf 1704} (2017) 076
  [arXiv:1612.06548 [hep-ph]].
  
\bibitem{Biedermann:2017bss} 
  B.~Biedermann, A.~Denner and M.~Pellen,
  JHEP {\bf 1710}, 124 (2017)
  [arXiv:1708.00268 [hep-ph]].

\bibitem{Frederix:2017wme} 
  R.~Frederix, D.~Pagani and M.~Zaro,
  arXiv:1711.02116 [hep-ph].
  
\bibitem{Czakon:2017wor}
  M.~Czakon, D.~Heymes, A.~Mitov, D.~Pagani, I.~Tsinikos and M.~Zaro,
  arXiv:1705.04105 [hep-ph].
 
\bibitem{Aad:2015noh} 
  G.~Aad {\it et al.} [ATLAS Collaboration],
  Eur.\ Phys.\ J.\ C {\bf 76}, no. 2, 87 (2016)
  [arXiv:1509.02358 [hep-ex]].
  
\bibitem{Khachatryan:2015oga} 
  V.~Khachatryan {\it et al.} [CMS Collaboration],
  Phys.\ Lett.\ B {\bf 757}, 154 (2016)
  [arXiv:1507.03119 [hep-ex]].

\bibitem{Butterworth:2015oua}
  J.~Butterworth {\it et al.},
  J.\ Phys.\ G {\bf 43} (2016) 023001
  [arXiv:1510.03865 [hep-ph]].

\bibitem{Ball:2014uwa}
  R.~D.~Ball {\it et al.} [NNPDF Collaboration],
  JHEP {\bf 1504} (2015) 040
  [arXiv:1410.8849 [hep-ph]].

\bibitem{Harland-Lang:2014zoa}
  L.~A.~Harland-Lang, A.~D.~Martin, P.~Motylinski and R.~S.~Thorne,
  Eur.\ Phys.\ J.\ C {\bf 75} (2015) no.5,  204
  [arXiv:1412.3989 [hep-ph]].

\bibitem{Dulat:2015mca}
  S.~Dulat {\it et al.},
  Phys.\ Rev.\ D {\bf 93} (2016) no.3,  033006
  [arXiv:1506.07443 [hep-ph]].

\bibitem{Manohar:2016nzj}
  A.~Manohar, P.~Nason, G.~P.~Salam and G.~Zanderighi,
  Phys.\ Rev.\ Lett.\  {\bf 117} (2016) no.24,  242002
  [arXiv:1607.04266 [hep-ph]].

\bibitem{Manohar:2017eqh}
  A.~V.~Manohar, P.~Nason, G.~P.~Salam and G.~Zanderighi,
  arXiv:1708.01256 [hep-ph].

\bibitem{Czakon:2016dgf}
  M.~Czakon, D.~Heymes and A.~Mitov,
  JHEP {\bf 1704} (2017) 071
  [arXiv:1606.03350 [hep-ph]].

\bibitem{Czakon:2015owf}
  M.~Czakon, D.~Heymes and A.~Mitov,
  Phys.\ Rev.\ Lett.\  {\bf 116} (2016) no.8,  082003
  [arXiv:1511.00549 [hep-ph]].

\bibitem{Alwall:2014hca}
  J.~Alwall {\it et al.},
  JHEP {\bf 1407} (2014) 079
  [arXiv:1405.0301 [hep-ph]].

\bibitem{Frixione:2015zaa}
  S.~Frixione, V.~Hirschi, D.~Pagani, H.-S.~Shao and M.~Zaro,
  JHEP {\bf 1506} (2015) 184
  [arXiv:1504.03446 [hep-ph]].

\bibitem{Pagani:2016caq}
  D.~Pagani, I.~Tsinikos and M.~Zaro,
  Eur.\ Phys.\ J.\ C {\bf 76} (2016) no.9,  479
  [arXiv:1606.01915 [hep-ph]].

\bibitem{Martin:2009iq}
  A.~D.~Martin, W.~J.~Stirling, R.~S.~Thorne and G.~Watt,
  Eur.\ Phys.\ J.\ C {\bf 63} (2009) 189
  [arXiv:0901.0002 [hep-ph]].

\bibitem{Czakon:2011xx}
  M.~Czakon and A.~Mitov,
  Comput.\ Phys.\ Commun.\  {\bf 185} (2014) 2930
  [arXiv:1112.5675 [hep-ph]].

\bibitem{Sirunyan:2017lvd}
  M.~Aaboud {\it et al.} [ATLAS and CMS Collaborations],
  [arXiv:1709.05327 [hep-ex]].

\bibitem{web-based-results}
 Repository with results and additional plots of NNLO QCD + EW $t\t$ distributions and asymmetries: 
 \url{http://www.precision.hep.phy.cam.ac.uk/results/ttbar-nnloqcd-nloew/}
                   
\end{thebibliography}
\end{document}